\documentclass[]{elsarticle}
\usepackage{graphicx}
\begin{document}
\begin{frontmatter}

\title{Zen Puzzle Garden is {NP}-complete}

\author[mcr]{Robin Houston}
\author[lex]{Joseph White}
\author[mmu]{Martyn Amos}

\address[mcr]{mySociety Ltd., London, United Kingdom.}
\address[lex]{Lexaloffle Games, Wellington, New Zealand.}
\address[mmu]{School of Computing, Mathematics and Digital Technology, \\ Manchester Metropolitan University, United Kingdom.}


\end{frontmatter}

\section{Introduction}

\begin{figure}[!h]
   \begin{center}
   \includegraphics[scale=0.8]{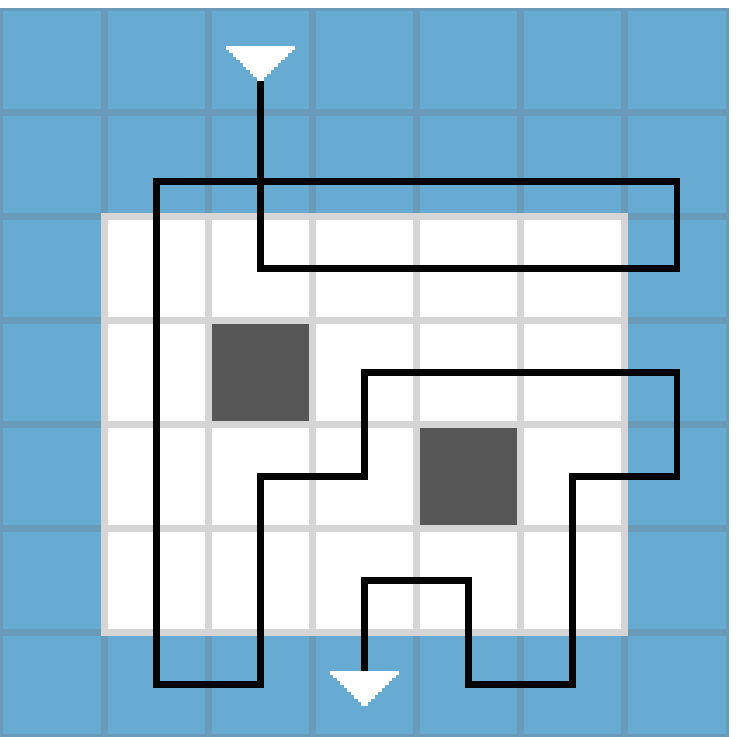}
       \caption{ZPG example garden, with solution. Sand is white, walkable squares are blue, and rocks are grey.}
       \label{garden}
       
\vspace{0.6cm}

   \includegraphics[scale=0.5]{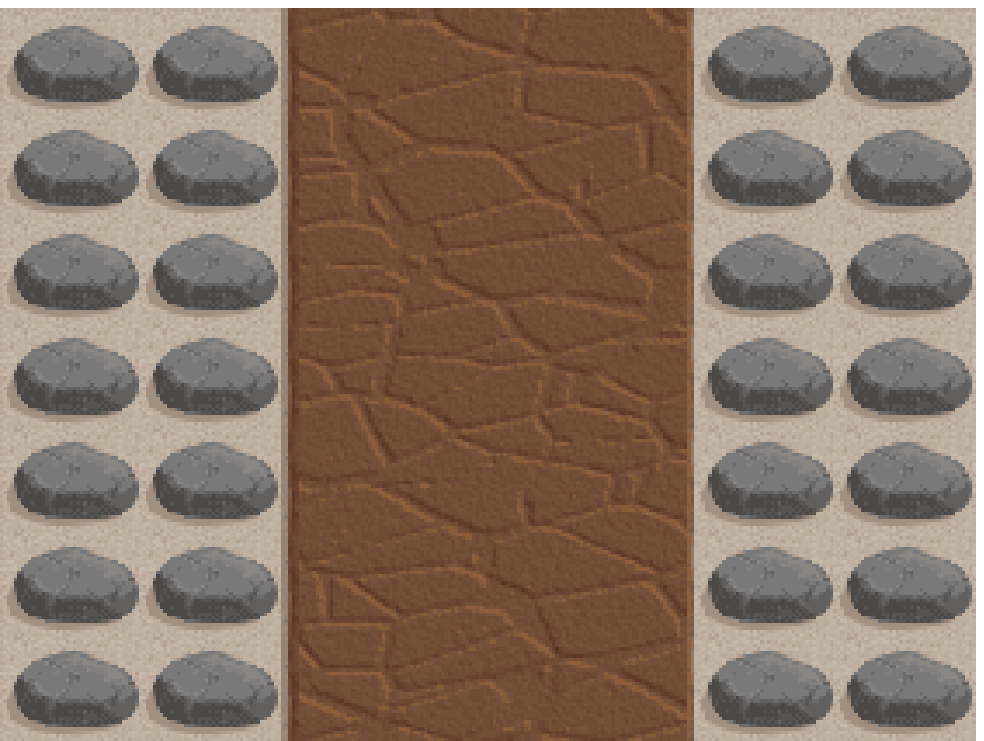}
    \includegraphics[scale=0.5]{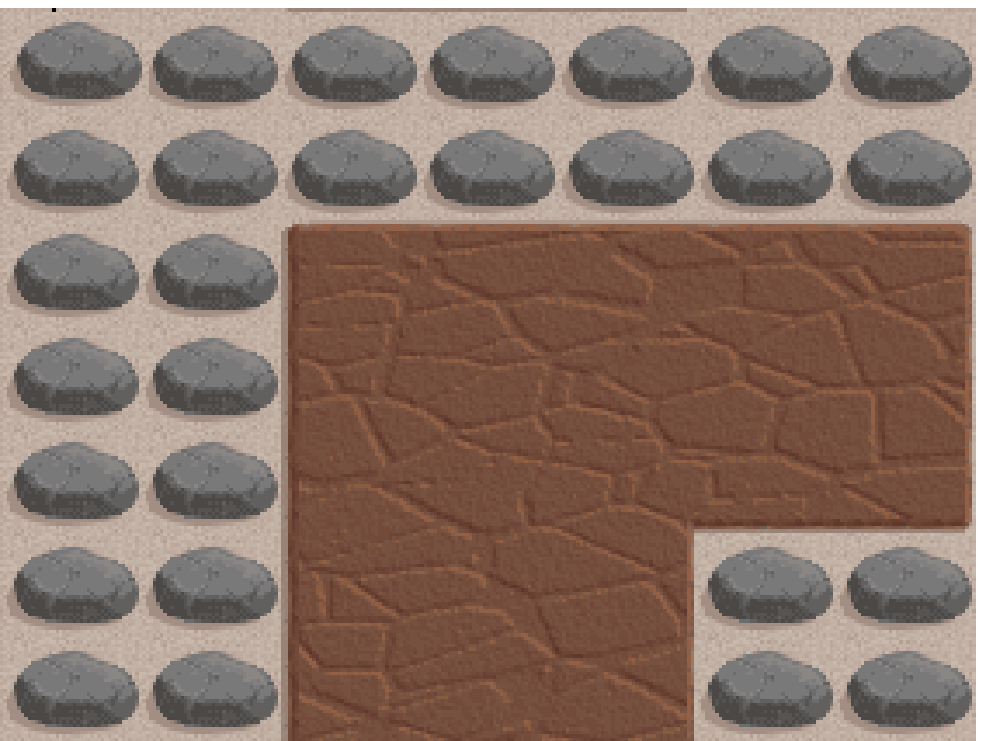}
       \caption{(Left) Straight edge and (Right) corner tiles, represented using actual game graphics.  Sand is coloured beige, and walkable squares are brown. Rocks are grey.}
       \label{straight-corner}

\vspace{0.6cm}

   \includegraphics[scale=0.5]{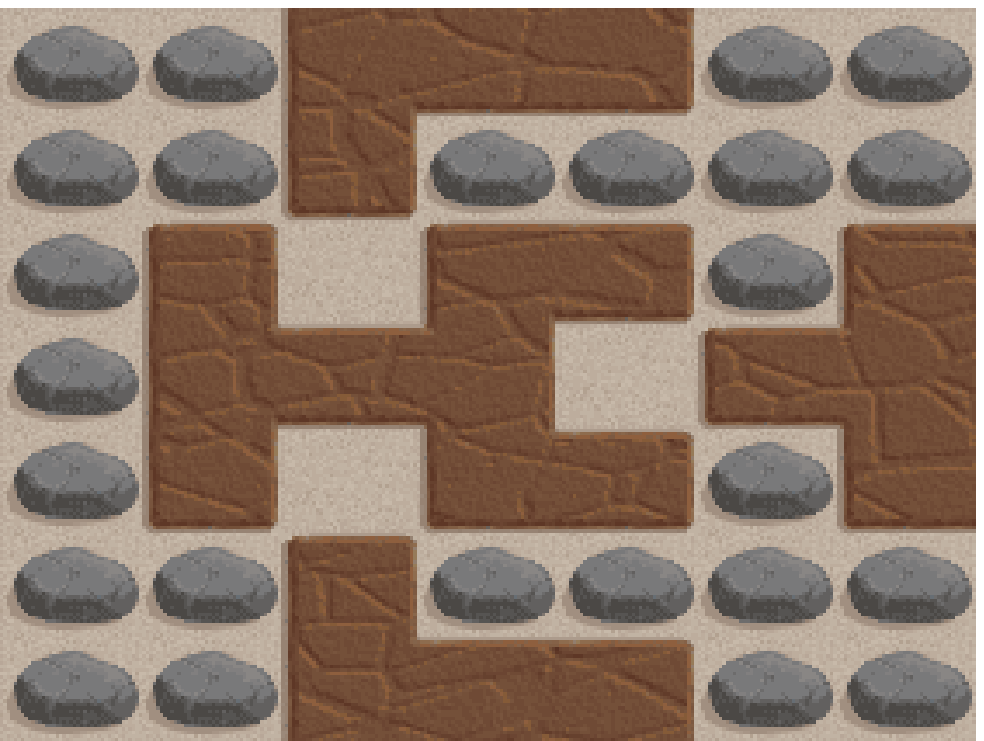}
    \includegraphics[scale=0.5]{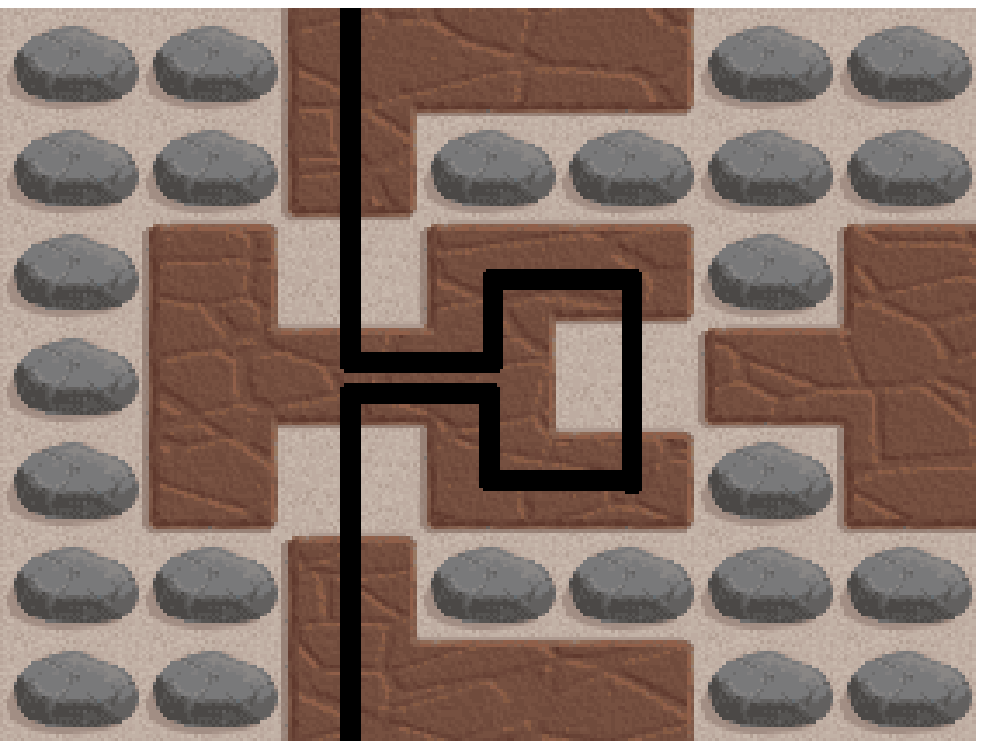}
       \caption{(Left) Node tile, and (Right) example path.}
       \label{node}
       
   \end{center}
 \end{figure}

Zen Puzzle Garden (ZPG) is a one-player puzzle game that takes place on a two-dimensional grid of squares, called a {\it garden} \cite{Lex}. Squares may be either {\it sand}, {\it rock}, or {\it walkable}. The objective of the game is to move a {\it monk} character around the garden, causing him to cover {\it all} sand squares.  The monk may move freely on walkable squares. When on a sand square the monk may only move within the von Neumann neighbourhood (i.e., no diagonal movements are allowed). Once moving on sand, the monk continues to move in a straight line until he encounters either a walkable square (in which case he moves onto it), a covered sand square or a rock (in both cases, he stops moving). The monk may not turn {\it while moving} on sand. An example garden is depicted in Figure ~\ref{garden}, along with a sample solution.

Although the problem has been studied experimentally \cite{ga}, until now no formal proof of its complexity has existed. We now demonstrate that deciding the solvability of ZPG is NP-complete. \\

\noindent {\bf Theorem 1:}
Deciding the solvability of a Zen Puzzle Garden instance is NP-complete. \\

Following \cite{pearl}, we construct a reduction from the following NP-complete problem \cite{garey}; given a cubic planar graph, does it contain a Hamiltonian circuit? That is, we construct ZPG gardens which correspond to arbitrary cubic planar graphs, and any garden we construct will have a solution iff the corresponding graph has a Hamiltonian circuit.

\newpage

\newpage
\section{Reduction}

To build a ZPG garden that corresponds to a cubic planar graph, $G$, we first draw the graph on a grid. This may always be done in such a way that the area of the grid is quadratic in the size of the graph \cite{grid}. We then convert the grid into a ZPG garden as follows. Each square of the grid is converted into a 7$\times$7 tile, made up of either rock, sand or walkable squares. Each tile is either a straight edge, a corner, or a node of the graph with three incident edges. The first two tiles are depicted in Figure ~\ref{straight-corner}, and these may be rotated as required.
 
The node tile is constructed in such a way that, whichever sand square is covered first, it is always possible to cover the remaining sand squares and come out either of the other sides (Figure ~\ref{node}).

\begin{figure}[]
   \begin{center}
   \includegraphics[scale=0.5]{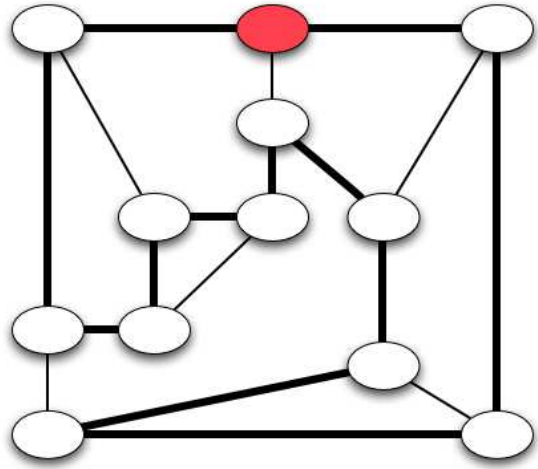}
    \includegraphics[scale=0.45]{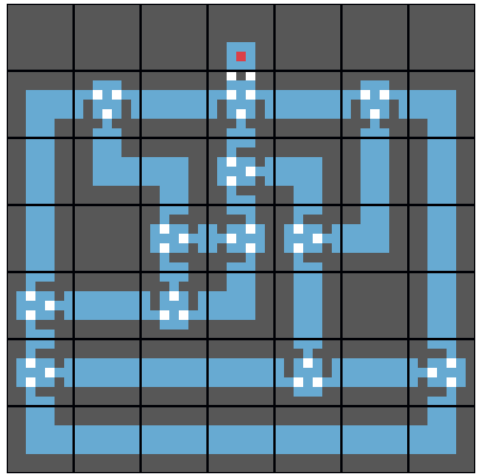}\\
    \vspace{0.6cm}
     \includegraphics[scale=0.5]{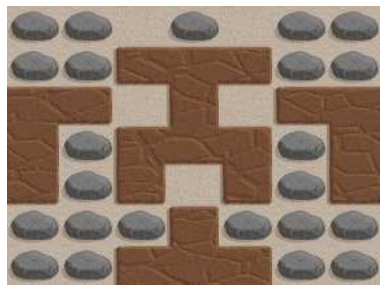}
     \vspace{0.3cm}
       \caption{(Top) Cubic planar graph. (Middle) ZPG representation. As before, sand squares are coloured white, walkable squares are blue and rocks are grey. (Bottom) Gateway tile.}
       \label{graph}
   \end{center}
 \end{figure}

In Figure ~\ref{graph} we show a cubic planar graph, along with its ZPG representation. The start node is coloured in red, and the Hamiltonian cycle is shown in bold. 

In the ZPG representation, the start node is encoded as a two-tile complex, occupying positions (3,6) and (3,5). The top tile is made entirely of walkable squares, where the monk is initially placed, and the lower tile is a single ``gateway" tile, depicted in Figure ~\ref{graph} (bottom). This tile is a modified node tile with two of the upper-most rocks converted to sand. The first sand square covered must be one of these two, and the last sand square visited must be the other. This facilitates exit from and return to the start node. Any solution must pass through all other nodes only once, so any solution to the garden gives a Hamiltonian circuit though the graph.

We know from \cite{grid} that a planar graph on $n$ nodes can be realised using a grid of area $O(n^2)$, and that this grid realisation can be computed in time $O(n \log n)$, so the mapping from planar graph to ZPG garden is polynomial. Given a solution, it can easily be checked in polynomial time, since a solution is never bigger than the garden it solves. This completes the proof that ZPG is NP-complete.

\end{document}